\documentclass[%
 reprint,
 superscriptaddress,
preprintnumbers,
 amsmath,amssymb,
 aps,
 pre,
 showpacs,
 floatfix,
]{revtex4-1}
\usepackage{graphicx}
\usepackage{dcolumn}
\usepackage{bm}
\usepackage{time}
\usepackage{subfigure}
\newcommand\Tstrut{\rule{0pt}{2.9ex}}         
\newcommand{\expect}[1]{\langle \, #1 \, \rangle}
\newcommand{\dd}[1]{\mathrm{d} #1 \,} 
\newcommand{\DD}[1]{\mathcal{D} #1 \,} 
\newcommand{\sech}[1]{\mathrm{sech} #1 }      
\newcommand{\bphi}{\bm{\phi}}
\newcommand{\bphistar}{\bm{\phi}^{\star}}
\newcommand{\bPhi}{\bm{\Phi}}
\newcommand{\bPhistar}{\bm{\Phi}^{\star}}
\newcommand{\boldeta}{\bm{\eta}}
\newcommand{\btheta}{\bm{\theta}}
\newcommand{\bG}{\mathbf{G}}
\newcommand{\bD}{\mathbf{D}}
\newcommand{\bF}{\mathbf{F}}
\newcommand{\bJ}{\mathbf{J}}
\newcommand{\bM}{\mathbf{M}}
\newcommand{\rmi}{{\mathrm i}}
\newcommand{\rmd}{{\mathrm d}}
\newcommand{\rme}{{\mathrm e}}

\newcommand{\be}{\begin{equation}}
\newcommand{\bq}{\begin{equation}}
\newcommand{\ee}{\end{equation}}
\newcommand{\eq}{\end{equation}}
\newcommand{\bea}{\begin{eqnarray}}
\newcommand{\eea}{\end{eqnarray}}
\newcommand{\ba}{\begin{eqnarray}}
\newcommand{\ea}{\end{eqnarray}}
\newcommand{\tu} {\tilde{u}} 
\newcommand{\tv} {\tilde{v}} 
\newcommand{\tl} {\tilde{l}} 
\usepackage{hyperref}
\hypersetup{
    pdfnewwindow=true,      
    colorlinks=true,        
    linkcolor=red,          
    citecolor=blue,         
    filecolor=green,        
    urlcolor=cyan           
}
\begin{document}
\preprint{LA-UR-19-30114}
%
%
\title{ Composite Molecules and Decoupling in Reaction Diffusion Models}
\author{John F. Dawson}
\email{john.dawson@unh.edu}
\affiliation{Department of Physics,
   University of New Hampshire,
   Durham, NH 03824, USA}   
\author{Fred Cooper} 
\email{cooper@santafe.edu}
\affiliation{The Santa Fe Institute, 
   1399 Hyde Park Road, 
   Santa Fe, NM 87501, USA}
\affiliation{Theoretical Division,
   Los Alamos National Laboratory,
   Los Alamos, NM 87545}
\author{Bogdan Mihaila}
\email{bmihaila05@gmail.com}
\affiliation{Physics Division, National Science Foundation}
\date{\today, \now \ PDT}
%
%
\begin{abstract}
The  Gray-Scott model can be thought of as an effective theory at large spatiotemporal scales coming from a more fundamental theory valid at shorter spatiotemporal scales.  The more fundamental theory includes a composite molecule which is trilinear in the molecules of the Gray-Scott model as was shown in the recent derivation of the Gray-Scott model from the master equation.  Here we show that at a classical level, ignoring the fluctuations describable in a Langevin description, the late time dynamics of the more fundamental theory leads to the same pattern formation as found in the Gray-Scott model with suitable choices of the parameters describing the diffusion of the composite molecule.
\end{abstract}
%
%
\maketitle
%
%
\section{\label{s:intro}Introduction}

Decoupling phenomena are well studied in quantum field theory (see for example \cite{PhysRevD.24.481}), where it can be shown that heavy degrees of freedom that are present in a theory appropriate at high energies and  short distances can be integrated out.  This then leads to local two-body interactions in the effective theory valid at low energy or equivalently at large temporal and spatial scales.  This happens for example in the Glashow-Weinberg-Salam gauge theory of weak interactions \cite{Glashow} 
\cite{Weinberg-Weak}  mediated by a heavy $W$ boson.  At low energies, the Glashow-Weinberg-Salam theory is described by Fermi's theory of local contact 4-Fermi interactions.  It has been shown in the context of an auxiliary field loop expansion that certain bilinear combinations in the 4-Fermion model, are the low energy analogues of the heavy boson as well as the scalar mesons in the weak interaction theory \cite{PhysRevLett.40.1620}, and they mediate the dynamics of the weakly interacting fermions.  The only difference occurring in the exact Schwinger-Dyson equation for the inverse vector meson propagators of the 4-Fermi theory was the replacement of the free vector meson inverse propagator of the  fundamental theory by a contact term. After renormalization of both theories the low energy properties of both theories become identical. 

A few years ago \cite{PhysRevLett.111.044101} we investigated the decoupling hypothesis in the context of the Gray-Scott model \cite{r:Gray:1983fk} and introduced two possible candidate dynamical states made up of two molecules of the original model.   Without having more intuition at that time, we showed that the decoupling hypothesis worked with these assumptions.  A major flaw in that analysis was any evidence that these particular composite states were important in the underlying chemistry.  Subsequent theoretical studies \cite{PhysRevE.88.042926} have identified the correct composite chemicals that play a part in the chemical reactions of the Gray-Scott model \cite{r:Gray:1983fk} when treated in a more fundamental fashion  which include fluctuations. In that study it was shown that  that the chemical reaction dynamics resembled the dynamics found in the quantum theory of weak interaction physics in mean-field approximation.  

The purpose of this article is to remedy this shortcoming of our previous analysis \cite{PhysRevLett.111.044101}. In our previous paper on decoupling \cite{PhysRevLett.111.044101} we assumed that the analogue of the intermediate boson, which can be thought of as a particle anti-particle composite particle, were the composite states $UV$ and $V^2$.  More recently \cite{PhysRevE.88.042926}, we were able to derive the Gray-Scott model from a master equation and therefore include fluctuations arising from the stochastic nature of the underlying scattering processes.  By doing this we were able to show that the fluctuations induced $N \rightarrow N$ scattering processes all of which proceed through the propagator of the composite state $UV^2$.  The exact inverse propagator  of the  composite state $U+2 V$ consisted of the contact 6 point interaction followed by the infinite  sum of 3 particle multiple re-scattering graphs (two loop graphs made of one $U$ propagator and two $V$ propagators).  On the other hand the composites $U+U$ and $U+V$ entered into the induced interaction ``tree diagrams'' only as single re-scattering processes and they did not turn into  fully  propagating entities (in the language of field theory). 
Thus for the chemistry of the Gray-Scott model, where the basic ``local'' form of the interaction is a 6 boson interaction  (three molecules entering and three leaving) as opposed to the 4-fermi interaction of weak interaction physics, the composite molecule $W \equiv UV^2$ is the analogue of the intermediate boson of the theory of weak interactions between electrons, muons and neutrinos \cite{PhysRevLett.40.1620}.  Also we were able to show that the full theory including the intrinsic fluctuations was describable in terms of the original Gray Scott model by having multiplicative noise terms driving the reaction diffusion equation for the $U$ and $V$ molecules.  An important feature of this description was the fact that the composite field $w(x)$, corresponding to the composite molecule $W = UV^2$ is  also  driven by a multiplicative noise term.  We repeat the derivation of \cite{PhysRevE.88.042926},  in Section~\ref{s:action}.  Thus to extend the analogy with the Weak interaction Fundamental theory, we consider promoting the field $W$ to a fundamental composite chemical agent, and show here that the pattern formation obtained at large scales is almost identical to what is found in the original Gray Scott model.

%
%
\section{\label{s:GrayScott}The Gray-Scott model}

The Gray-Scott model involves two species $U$ and $V$ that undergo the chemical reactions:
\begin{equation}\label{GS.e:1}
   U + 2 V \xrightarrow{\lambda} 3 V \>,
   \>\>
   U \xrightarrow{\nu} Q \>,
   \>\>
   V \xrightarrow{\mu} P \>,
   \>\>
   \xrightarrow{f} U \>.
\end{equation}
There is a cubic autocatalytic step for $V$ at rate $\lambda$, and decay reactions at rates $\mu, \nu$ that transform $V$ and $U$ into inert products $P$ and $Q$. Finally, $U$ is fed into the system at a rate $f$. The phenomenological approach to study the dynamics of such systems utilizes the law of mass action and allows us to interpret the chemical reaction $U + 2 V \xrightarrow{\lambda} 3 V$ as having the terms $\pm \lambda u v^2$ in its reaction kinetics.  Following this and including diffusion as a first approximation to molecular motion, the equations that describe the kinetics of the system are
\begin{subequations}\label{GSequations}
\begin{align}
   \bigl [ \,
      \partial_t - D_{u} \, \nabla^2 + \nu \,
   \bigr ] \, u
   +
   \lambda \, u \, v^2
   &=
   f \>,
   \label{GS.e:2a}  \\
   \bigl [ \,
      \partial_t - D_{v} \, \nabla^2 + \mu \,
   \bigr ] \, v
   -
   \lambda \, u \, v^2
   &=
   0 \>.
   \label{GS.e:2b}
\end{align}
\end{subequations}
Here $u(x)$ and $v(x)$ are fields in a $d+1$ dimensional space $x \equiv (\mathbf x, t)$, and represent the concentrations of the chemical species $U$ and $V$.  $D_v$ and $D_u$ are diffusion constants for species $V$ and $U$ respectively. 
The red steady state solutions are given by
\begin{equation}\label{SC.e:red}
   u_0 = \frac{f}{\nu}
   \,~~
   v_0 = 0 \>.
\end{equation}

%
%
\section{\label{s:action}Derivation of the action}

The many-body formulation of many types of reaction and diffusion processes is discussed thoroughly in the literature \cite{r:Vollmayr-Lee:1994nr,r:Tauber:2005rc,r:Cardy:1999fk,r:Tauber:2014bf,r:Baez:2012jk}.  The standard procedure for obtaining a path integral is to start from the master equation for the reaction and diffusion processes, develop a number algebra with annihilation and creation operators using a Hilbert space, define a conserving state vector $\Psi(t)$ and a Schr{\"o}dinger-like equation, pass over to a continuum description, and then write a path integral for the generating functional.  This procedure yields a Doi-shifted \cite{0305-4470-9-9-009} path integral of the form
\begin{equation}\label{A.e:1}
   Z
   =
   \iint \DD{\bphistar} \DD{\bphi} \exp\{ - S[\bphistar,\bphi]\, \} \>,
\end{equation}
where
\begin{align}
   S[\bphistar,\bphi]
   &=
   \int \dd{x} 
   \bigl \{ \,
      \phi^{\star}_u(x) \, \bigl [\, G^{-1}_u(x) \, \phi_u(x) - f \, \bigr ]
      \notag \\[-4pt]
      & \qquad
      +
      \phi^{\star}_v(x) \, G^{-1}_v(x) \, \phi_v(x)
      \label{A.e:2} \\
      & \qquad
      +
      \sigma(x) \, (\, \phi^{\star}_u(x) - \phi^{\star}_v(x) \, ) \,
      ( \, 1 + \phi^{\star}_v(x) \,)^2 \,
   \bigr \} \>.
   \notag
\end{align}
Here we have set $\sigma(x) = \lambda \, \phi_u(x) \phi_v^2(x)$, and put
\begin{subequations}\label{A.e:3}
\begin{align}
   G^{-1}_u(x)
   &=
   \partial_t - D_{u} \, \nabla^2 + \nu \>,
   \label{A.e:3a} \\
   G^{-1}_v(x)
   &=
   \partial_t - D_{v} \, \nabla^2 + \mu \>.
   \label{A.e:3b}
\end{align}
\end{subequations}
The fields $\phi_{u,v}(x)$ and $\phi^{\star}_{u,v}(x)$ are eigenvalues of the creation and annihilation operators respectively, and are \emph{not} complex conjugates of each other.  The action in \eqref{A.e:2} has a cubic term in $\phi^{\star}_v$.     At the cost of adding auxiliary fields  $\phi_w$ and $\phi_w^{\star}$ which are quadratic in  $\phi^\star_v$ we can render the Lagrangian quadratic in the starred fields, which then allows the theory to have a Langevin equation description. This is accomplished by introducing a representation of unity, 
\begin{align}\label{A.e:4}
   &1
   =
   \int \DD{\phi_w^{\star}} \delta(\, \phi_w^{\star} - \phi^{\star\,2}_v \,)
   \\
   &=
   \iint\! \DD{\phi_w^{\star}} \DD{\phi_w} 
   \exp \Bigl \{ 
      - \!\int\! \dd{x} \phi_w(x) \, 
      (\,  \phi_w^{\star}(x) - \phi^{\star\,2}_v(x) \, ) \, 
        \Bigr \}
   \notag
\end{align}
into the path integral.  Using the notation 
\begin{equation}\label{A.e:5}
   \bPhi 
   = 
   \begin{pmatrix} \phi_v \\ \phi_u \\ \phi_w \end{pmatrix}
    \>,
   \qquad
   \bPhistar 
   =
   \begin{pmatrix} \phi_v^{\star} \\ \phi_u^{\star} \\ \phi_w^{\star} \end{pmatrix} 
\end{equation}
the path integral is written as
\begin{equation}\label{A.e:6}
   Z
   =
   \iint \DD{\bPhistar} \DD{\bPhi} \exp\{ - S[\bPhistar,\bPhi]\, \} \>,
\end{equation}
where now the action is given by
\begin{align}\label{A.e:7}
   S[\bPhistar,\bPhi]
   &=
   \int \dd{x} \bPhistar{}^T(x) \cdot [ \, \bG^{-1}(x) \cdot \bPhi(x) - \bF(x) \, ]
   \\
   & \qquad
   -
   \frac{1}{2}
   \int \dd{x} \bPhistar{}^T(x) \cdot \bD[\bPhi](x) \cdot \bPhistar(x) \>,
   \notag
\end{align}
where
\begin{equation}\label{A.e:8}
   \bG^{-1}(x)
   =
   \begin{pmatrix} 
      G^{-1}_v(x) & 0 & 0 \\
      0 & G^{-1}_u(x) & 0 \\
      0 & 0 & 0
   \end{pmatrix} \>,
   \>
   \bF(x) 
   = 
   \begin{pmatrix} 0 \\ f \\ 0 \end{pmatrix} \>,
\end{equation}
and the noise correlation matrix is given by
\begin{equation}\label{A.e:10}
   \bD[\bPhi](x)
   =
   \begin{pmatrix}
      2 \phi_w(x) & -2 \sigma(x) & \sigma(x) \\
      -2 \sigma(x) & 0 & -\sigma(x) \\
      \sigma(x) & -\sigma(x) & 0 
   \end{pmatrix} \>.
\end{equation}
Now using the identity,
\begin{align}
   &\sqrt{\det{\bD}}
   \exp \Bigl \{
      \frac{1}{2}
      \iint \dd{x} \dd{x'} \,
      \bPhistar(x) \cdot \bD(x,x') \cdot \bPhistar(x') \Bigr \}
   \notag \\
   & \quad
   =
   \int \! \DD{\boldeta} \,
   \exp \Bigl \{ 
      - \frac{1}{2}
      \iint \dd{x} \dd{x'} \,
      \boldeta(x) \cdot \bD^{-1}(x,x') \cdot \boldeta(x)
      \notag \\
      & \qquad
      +
      \int \dd{x} \, \bPhistar(x) \cdot \boldeta(x) 
         \Bigr \} \>,
   \label{A.e:11}
\end{align}
where
\begin{equation}\label{A.e:12}
   \boldeta(x)
   =
   (\, \eta_v(x), \eta_u(x), \eta_w(x) \,) \>,
\end{equation}
the path integral \eqref{A.e:6} becomes
\begin{equation}\label{A.e:13}
   Z
   =
   \int \!\DD{\boldeta} \!\int \!\DD{\bPhi} P[\,\bPhi,\boldeta\,] 
   \!\int \!\DD{\bPhistar}  
   \rme^{ - S[\bPhistar,\bPhi,\boldeta]\, } \>,
\end{equation}
where
\begin{align}
   &P[\, \boldeta,\bPhi \,]
   \label{A.e:14} \\
   & \>
   =
   \mathcal{N}
   \exp \Bigl \{ 
      - \frac{1}{2}
      \iint \dd{x} \dd{x'} 
      \boldeta(x) \cdot \bD^{-1}[\bPhi](x,x') \cdot \boldeta(x) \Bigr \} \>, 
   \notag 
\end{align}
and the action is now given by
\begin{equation}\label{A.e:15}
   S[\bPhistar,\bPhi,\boldeta]
   =
   \int \dd{x} \bPhistar(x) \cdot [ \, \bG^{-1}(x) \cdot \bPhi(x) - \bJ(x) \, ] \>,
\end{equation}
with
\begin{equation}\label{A.e:16}
   \bJ(x)
   =
   (\, f + \eta_u(x), \eta_v(x), \eta_w(x) \, ) \>.
\end{equation}
Integrating \eqref{A.e:13} over $\bPhistar$ then yields,
\begin{equation}\label{A.e:17}
   Z
   =
   \int \!\DD{\boldeta} \!\int \!\DD{\bPhi} P[\,\bPhi,\boldeta\,] 
   \delta \bigl [\, \bG^{-1}(x) \cdot \bPhi(x) - \bJ(x) \, \bigr ] \>,
\end{equation}
Finally, the $\boldeta(x)$ noise functions can be related to white noise sources by performing a Cholesky decomposition of the  correlation matrix \eqref{A.e:10}:
\begin{equation}\label{A.e:18}
   \bD(x)
   =
   \bM^{T}(x) \cdot \bM(x) \>, 
\end{equation}
where one possible choice is
\begin{equation}\label{A.e:19}
   \bM^{T}
   =
   \frac{1}{\sqrt{2 \phi_w}}
   \begin{pmatrix}
      2 \phi_w & 0 & 0 \\
      -2 \sigma & -2i \sigma & 0 \\
      \sigma & i ( \sigma - \phi_w ) & i \sqrt{\phi_w ( 2 \sigma - \phi_w ) }
   \end{pmatrix} \>,
\end{equation}
and writing
\begin{equation}\label{A.e:20}
   \boldeta^{T} \cdot \bD^{-1} \cdot \boldeta
   =
   \boldeta^{T} \cdot \bM^{-1} \cdot [ \, \bM^{T} \,]^{-1} \cdot \boldeta
   =
   \btheta^{T} \cdot \btheta \>,
\end{equation}
where we have put $\btheta = (\, \theta_1,\theta_2,\theta_3 \,)$.  So $\boldeta = \bM^T \cdot \btheta$, which gives complex noise functions:
\begin{subequations}\label{A.e:21}
\begin{align}
   \eta_v
   &=
   \sqrt{2\phi_w} \, \theta_1 \>,
   \label{A.e:21a} \\
   \eta_u
   &=
   - \frac{2 \sigma}{\sqrt{2\phi_w}} \, ( \theta_1 + i \theta_2 )
   \label{A.e:21b} \\
   \eta_{w}
   &=
   \frac{\sigma}{\sqrt{2\phi_w}} \, \theta_1 
   + 
   i \frac{\sigma - \phi_w}{\sqrt{2\phi_w}} \, \theta_2
   +
   i \sqrt{\sigma - \frac{\phi_w}{2}}  \, \theta_3
   \label{A.e:21c}
\end{align}
\end{subequations}
The factorization here is not unique.  The path integral \eqref{A.e:17} then becomes
\begin{equation}\label{A.e:22}
   Z
   =
   \int \!\DD{\btheta} P[\,\btheta\,] \!\int \!\DD{\bPhi}
   \delta \bigl [\, \bG^{-1}(x) \cdot \bPhi(x) - \bJ(x) \, \bigr ] \>,
\end{equation}
where $P[\,\btheta\,]$ is the white noise probability distribution,
\begin{equation}\label{A.e:23}
   P[\,\btheta\,]
   =
   \mathcal{N} \, 
   \exp \Bigl \{ - \frac{1}{2} \int\! \dd{\btheta} 
   \btheta^T \cdot \btheta \Bigr \} \>. 
\end{equation}
The path integral has value only when $\bPhi(x)$ satisfies the Langevin equations,
\begin{equation}\label{A.e:24}
   \bG^{-1}(x) \cdot \bPhi(x) = \bJ(x) \>,
\end{equation}
or in component form, when
\begin{subequations}\label{A.e:25}
\begin{align}
   [\, \partial_t - D_{u} \, \nabla^2 + \nu \,] \, \phi_u(x)
   +
   \sigma(x)
   &=
   f + \eta_u(x) \>,
   \label{A.e:25a}  \\
   [\, \partial_t - D_{v} \, \nabla^2 + \mu \,] \, \phi_v(x)
   -
   \sigma(x) 
   &=
   \eta_v(x) \>,
   \label{A.e:25b} \\
   \phi_w(x) - 2 \, \sigma(x)
   &=
   \eta_w(x) \>,
   \label{A.e:25c}
\end{align}
\end{subequations}
where $\boldeta$ is given in \eqref{A.e:21} and $ \sigma(x)$ is shorthand for $\lambda \phi_u \phi_v^2$ The asymmetry of these equations is seen in the fact that $\phi_w$ is a constraint field, just as the combination vector and axial vector currents in the original 4-Fermi theory of weak interactions were constrained fields.  

Note that the fields $\bphi$ in \eqref{A.e:25} are complex.  The connection with the classical level equations~\eqref{GSequations} are made by the identification that the noise-averaged fields correspond to real densities.
When we set the noise to zero we obtain the equations of the Gray-Scott model \eqref{GSequations}, when we identify $\phi_u$ with $u$, $\phi_v$ with $v$. 
\begin{subequations}\label{A.e:27}
\begin{align}
   [\, \partial_t - D_{u} \, \nabla^2 + \nu \,] \, u(x)
   +
   \lambda \, u(x) \, v^2(x)
   &=
   f \>,
   \label{A.e:27a}  \\
   [\, \partial_t - D_{v} \, \nabla^2 + \mu \,] \, v(x)
   -
   \lambda \, u(x) \, v^2(x) 
   &=
   0 \>,
   \label{A.e:27b} \\
   w(x) - 2 \, \lambda \, u(x) \, v^2(x)
   &=
   0 \>,
   \label{A.e:27c}
\end{align}
\end{subequations}
 We can rewrite these equations in the suggestive form:
\begin{subequations}\label{A.f:27}
\begin{align}
   [\, \partial_t - D_{u} \, \nabla^2 + \nu \,] \, u(x)
   +
  \frac{1}{2} w(x) 
   &=
   f \>,
   \label{A.f:27a}  \\
   [\, \partial_t - D_{v} \, \nabla^2 + \mu \,] \, v(x)
   -
 \frac{1}{2} w(x) 
   &=
   0 \>,
   \label{A.f:27b} \\
   w(x) - 2 \, \lambda \, u(x) \, v^2(x)
   &=
   0 \>,
   \label{A.f:27c}
\end{align}
\end{subequations}

%
%
\section{\label{s:GSphases}Phases of the Gray-Scott Model}

The regimes in parameter space where various stable spatiotemporal patterns have been observed have been mapped out in various investigations, both theoretical and numerical \cite{Pearson:1993aa,r:Mazin:1996cr,OSIPOV1996400}.  Static phases are found as solutions of Eqs.~\eqref{A.e:27},
\begin{subequations}\label{GS.e:F-1}
\begin{align}
   \nu \, u_0 + \lambda \, u_0^{\phantom 2} \, v_0^2 
   &=
   f \>,
   \label{GS.e:F-1a} \\
   \mu \, v_0 - \lambda \, u_0^{\phantom 2} \, v_0^2 
   &=
   0 \>,
   \label{GS.e:F-1b}
\end{align}
\end{subequations}
There are two sets of solutions of \eqref{GS.e:F-1} given by
\begin{subequations}\label{GS.e:F-2}
\begin{align}
   u_0 &= f / \nu \>,
   \qquad
   v_0 = 0 \>,
   \qquad
   \chi_0
   =
   0 \>,
   \label{GS.e:F-2a} \\
   u_0^{(\pm)}
   &=
   \frac{f \pm \sqrt{f^2 - f_{\text{min}}^2}}{2 \nu}
   =
   \frac{\mu}{\sqrt{\nu \lambda}} \,
   \rme^{\pm \theta } \>,   
   \label{GS.e:F-2b} \\
   v_0^{(\pm)}
   &=
   \frac{f \mp \sqrt{f^2 - f_{\text{min}}^2}}{2 \mu}
   =
   \frac{\nu}{\sqrt{\nu \lambda}} \,
   \rme^{\mp \theta } \>,
   \notag
\end{align}
\end{subequations}
where $f_{\text{min}}^2 = 4 \, \nu \mu^2 / \lambda$, and we have put
\begin{equation}\label{GS.e:F-3}
   f
   =
   f_{\text{min}} \, \cosh \theta \>.
\end{equation}
Solutions \eqref{GS.e:F-2a} are labeled ``red,'' and those in \eqref{GS.e:F-2b} ``blue.''  
One can perform a linear stability analysis of these states by inserting trial solutions of the form,
\begin{align}\label{GS.e:F-6}
   u(x) &= u_0 + \Delta u \, \rme^{\rmi ( \bm{k} \cdot \bm{x} - \omega t ) } \>,
   \\
   v(x) &= v_0 + \Delta v \, \rme^{\rmi ( \bm{k} \cdot \bm{x} - \omega t ) } \>,
   \notag
\end{align}
into \eqref{A.e:27} and examine the eigenvalues of the resulting Jacobian matrix.  The red state is always stable for any $(\bm{k},\omega)$ values, so we instead focus on the blue states.  For the blue states, the eigenvalues of the Jacobian matrix are given by
\begin{equation}\label{LR.e:17}
   \omega_k^{\pm}
   =
   - i \, B_k/2 \pm \sqrt{ C_k - (B_k/2)^2 } \>,
\end{equation}
where
\begin{subequations}\label{LR.e:16}
\begin{align}
   B_k
   &=
   (\, D_u + D_v \,) \, k^2 
   +
   \nu \, ( \rme^{\mp 2 \theta} + 1 )
   -
   \mu \>,
   \label{LR.e:18a} \\
   C_k
   &=
   D_u D_v \, k^4
   +
   [ \,
      D_v \, \nu \, ( \rme^{\mp 2 \theta} + 1 )
      -
      D_u \, \mu \, 
   ] \, k^2
   \label{LR.e:18b} \\
   & \qquad
   +
   \mu \, \nu \, ( \rme^{\mp 2 \theta} - 1 ) \>.
   \notag
\end{align}
\end{subequations}

%
%
\subsection{\label{GS.ss:HopfBi}Hopf bifurcation}

Homogeneous and oscillatory solutions are found for $\bm{k}=0$ when $B_0 = 0$.  The conditions when this happens is called the Hopf bifurcation, and ocurs when
\begin{subequations}\label{HB.e:1}
\begin{align}
   B_0
   &=
   \nu \, ( \rme^{\mp 2 \theta} + 1 ) - \mu
   =
   0 \>,
   \label{HB.e:1a} \\
   C_0
   &=
   \mu \, \nu \, ( \rme^{\mp 2 \theta} - 1 ) 
   >
   0 \>,
   \label{HB.e:1b}
\end{align}
\end{subequations}
so only the lower $(-)$ solutions give rise to a Hopf bifurcation.  The solution of \eqref{HB.e:1a} for the lower blue solutions is
\begin{equation}\label{HB.e:2}
   \rme^{2 \theta}
   =
   \frac{\mu - \nu}{\nu}
   \equiv
   \frac{\kappa}{\nu} \>,
\end{equation}
where we have set $\kappa = \mu - \nu > 0$.  The Hopf bifurcation occurs at a value of $f$ given by
\begin{equation}\label{HB.e:3}
   f_{\text{H}}
   =
   \frac{\mu^2}{\sqrt{ \lambda \, \kappa } } \>,
   \qquad
   \omega_{\text{H}}
   =
   \pm \sqrt{ \mu ( \kappa - \nu ) } \>.
\end{equation}
So $\kappa = \mu - \nu > \nu$ for an oscillations, otherwise the system is damped.  

%
%
\subsection{\label{GS.ss:TuringBi}Turing bifurcation}

\begin{figure}[t]
   \centering
   \includegraphics[width=0.9\columnwidth]{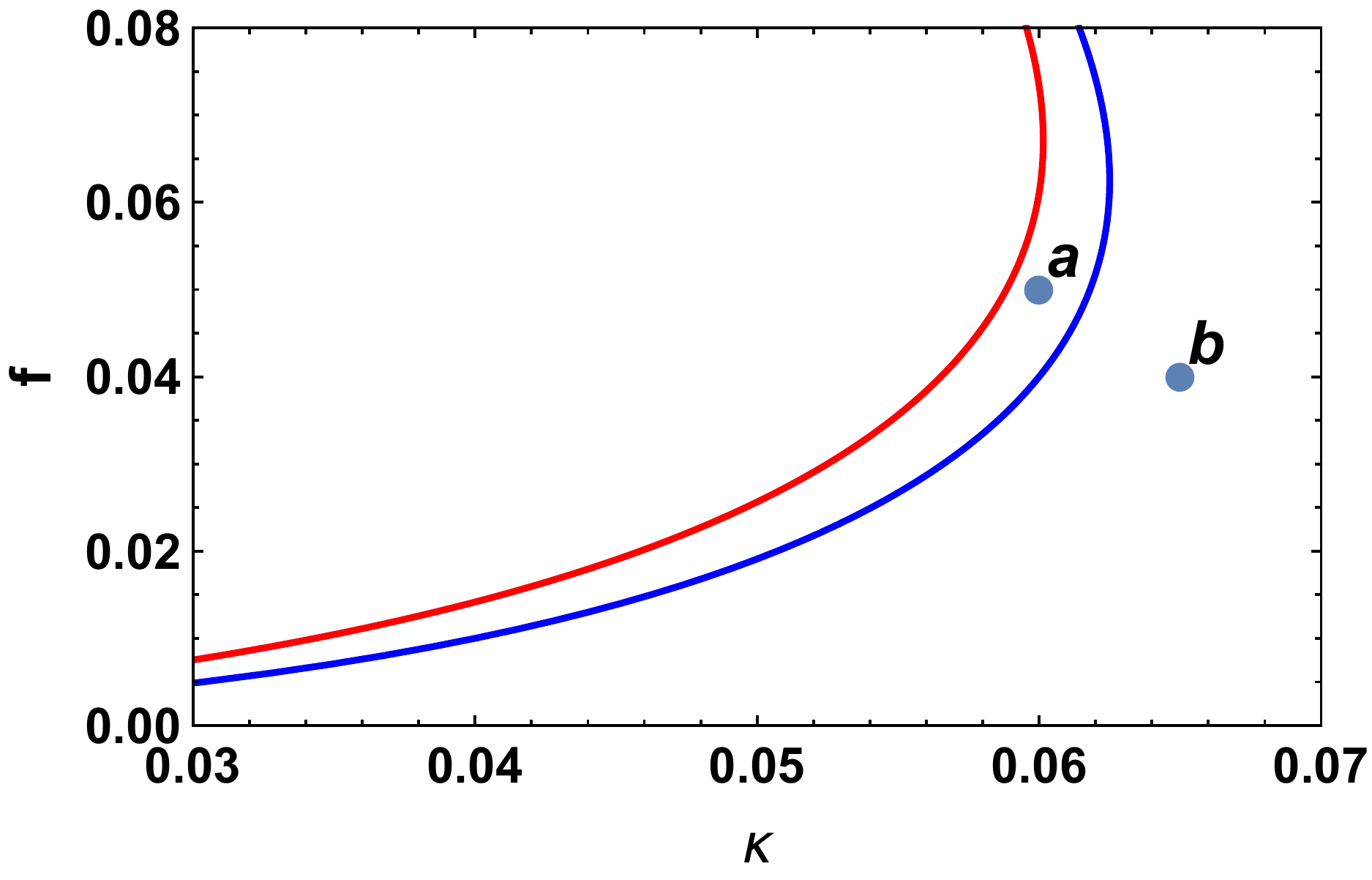}
   \caption{\label{fig:f-kappa-turing}Plot of $f$ \emph{vs} $\kappa$ for the
   case with $\nu = f$, $\mu = \nu + \kappa$, and with $\lambda = 1$ 
   and $\beta = 3$.  Turing instabilities occur for values of $f$
   and $\kappa$ in the region between the two curves.  Points $a$ and $b$
   label the values used in Video~\ref{vid:mov1} to produce ridge and spot 
   formation respectively.}
\end{figure}

The model also possesses a regime where the equilibrium solutions are unstable with respect to spatial perturbations---the so-called Turing instability.  For patterns to emerge, we must have
\begin{equation}\label{PF.e:m4-14}
   C_k \le 0 \>.
\end{equation}
The critical value is when $C_k = 0$.  As a function of $k^2$, $C_k$ is a parabolic curve with positive curvature,
\begin{subequations}\label{PF.e:m4-15}
\begin{align}
   \frac{ \partial C_k }{\partial k^2} 
   &=
   2 \, D_u D_v \, k^2
   +
   D_v \, \nu \, ( \rme^{\mp 2 \theta} + 1 )
   -
   D_u \, \mu \>,
   \label{PF.e:m4-15a} \\
   \frac{ \partial^2 C_k }{(\partial k^2)^2 } 
   &=
   2 \, D_u D_v
   > 0 \>.
   \label{PF.e:m4-15b}
\end{align}
\end{subequations}
The minimum of the curve is located at $k^2 = k_m^2$, so from \eqref{PF.e:m4-15a} given by
\begin{equation}\label{PF.e:m4-16}
   k_m^2
   =
   \frac{ D_u \, \mu - D_v \, \nu \, ( \rme^{\mp 2 \theta} + 1 )
        }{ 2 \, D_u D_v }
\end{equation}
The critical point is when the minimum of the parabolic curve touches the axis, that is when $k_m^2 = k_c^2$ with $C_{k_c} = 0$:
\begin{align}
   C_{k_c}
   &=
   D_u D_v \, k_c^4
   +
   [ \,
      D_v \, \nu \, ( \rme^{\mp 2 \theta} + 1 )
      -
      D_u \, \mu \, 
   ] \, k_c^2
   \label{PF.e:m4-17} \\
   & \qquad
   +
   \mu \, \nu \, ( \rme^{\mp 2 \theta} - 1 )
   =
   0 \>.
   \notag
\end{align}
Substitution of Eq.~\eqref{PF.e:m4-16} into \eqref{PF.e:m4-17}, we find:
\begin{equation}\label{PF.e:m4-17.2}
   D_u D_v \, k_c^4
   =
   \mu \, \nu \, ( \rme^{\mp 2\theta} - 1 )  \>,   
\end{equation}
or
\begin{equation}\label{PF.e:m4-18}
   k_c^2
   =
   \sqrt{ \frac{ \mu \, \nu \, ( \rme^{\mp 2\theta} - 1 ) }
               { D_u D_v } } \>.
\end{equation}
At the minimum value of $k_m = k_c$,
\begin{align}
   &
   [\, 
      D_u \, \mu 
      - 
      D_v \, \nu \, ( \rme^{\mp 2\theta} + 1 ) \,
   ]^2
   \label{PF.e:m4-20} \\
   & \hspace{4em} 
   =
   4 \, D_u D_v \, \mu \, \nu \, ( \rme^{\mp 2\theta} - 1 ) \>.
   \notag
\end{align}
Divide Eq.~\eqref{PF.e:m4-20} by $( D_v \, \nu )^2$, set $\beta = D_u / D_v$ and $\alpha = \beta \mu / \nu$, in which case \eqref{PF.e:m4-20} becomes
\begin{equation}\label{PF.e:m4-21}
   [\, \alpha - ( \rme^{\mp 2\theta} + 1 ) \,]^2
   =
   4 \, \alpha \, ( \rme^{\mp 2\theta} - 1 ) \>,
\end{equation}
which gives the equation:
\begin{equation}\label{PF.e:m4-23}
   \bigl [ \, \rme^{\mp 2 \theta} \, \bigr ]^2
   - 
   2 \, ( 3 \, \alpha - 1 ) \,
   \bigl [ \, \rme^{\mp 2 \theta} \, \bigr ]
   + 
   ( \alpha + 1 )^2 
   = 
   0 \>.
\end{equation}
The two solutions of this equation are:
\begin{equation}\label{PF.e:m4-24}
   \rme^{\mp2\theta_{\pm}}
   =
   ( 3 \, \alpha - 1 ) 
   \pm
   2 \sqrt{ 2 \alpha \, ( \alpha - 1 ) } \>.
\end{equation}
It turns out that the only solutions having Turing bifurcations are lower solutions of \eqref{GS.e:F-2b} given by
\begin{equation}\label{PF.e:24.1}
   u_0
   =
   \frac{\mu}{\sqrt{\nu \lambda}} \, \rme^{- \theta } \>,
   \qquad
   v_0
   =
   \frac{\nu}{\sqrt{\nu \lambda}} \, \rme^{+ \theta } \>,
\end{equation}
and of those the only one which gives $C_{k_c} = 0$ is the negative sign in \eqref{PF.e:m4-24}, that is
\begin{equation}\label{PF.e:m4-25}
   \rme^{2 \theta}
   =
   ( 3 \, \alpha - 1 ) 
   -
   2 \sqrt{ 2 \alpha \, ( \alpha - 1 ) } \>.
\end{equation}
Then, from Eq.\eqref{PF.e:m4-18}, the critical wavenumber $k_c$ is given by
\begin{align}\label{PF.e:m4-26}
   \frac{D_u D_v }{ \mu \, \nu } \, k_c^4
   =
   \rme^{ 2 \theta } - 1
   =
   \frac{1}{4 \alpha} \, [\, \alpha - ( \rme^{2\theta} + 1 ) \,]^2 \>,
   \notag
\end{align}
where we have used Eq.~\eqref{PF.e:m4-21}. 
A common set of parameters is to set $\lambda = 1$, $\beta = 3$, and to put $\nu = f$ and $\mu = \nu + \kappa$.  
For blue solutions to exist, $f \ge f_{\text{min}}$, so that for our case $\kappa \ge \sqrt{\lambda \nu} - \nu$.  
The maximum value of $\kappa$ for a given $\nu$ is found by solving
\begin{equation}\label{PF.e:27}
   \cosh^2\theta
   =
   \frac{\lambda \beta^2}{4 \nu \alpha^2} 
\end{equation}
for $\kappa$, where $\theta$ is given by \eqref{PF.e:m4-25}.  
The Turing instability region for this set of parameters is shown in Fig.~\ref{fig:f-kappa-turing}, where we have plotted $f$ \emph{vs} $\kappa$.  The region between the curves exhibits Turing instabilities.  This figure agrees with that in Mazin \cite{r:Mazin:1996cr} and in Hori \cite{Y.-Hori-and-S.-Hara:2012aa}.
%
%
%
%

%
%
\subsection{\label{ss:spots}Spot formation}

While the Turing stability results from infinitesimal spatial perturbations to the blue state, characterized by a global length-scale $q$, there is another regime where instabilities arise from \emph{large} excitation perturbations to the red state.  This regime was found by Osipov and Severtsev \cite{OSIPOV1996400} and Muratov and Osipov \cite{0305-4470-33-48-321}.  
First we rewrite Eqs.  \eqref{GSequations} as follows
\begin{subequations}\label{A.e:28}
\begin{align}
   \frac{\nu}{f} 
   \Bigl [\,
      \frac{1}{\nu} \partial_t - \frac{D_{u}}{\nu} \, \nabla^2 + 1 \,
   \Bigr ] \, u(x)
   +
   \frac{\lambda}{f} \, u(x) \, v^2(x)
   &=
   1 \>,
   \label{A.e:29a}  \\
   \mu \,
   \Bigl [\, 
      \frac{1}{\mu} \partial_t - \frac{D_{v}}{\mu} \, \nabla^2 + 1 \,
   \Bigr ] \, v(x)
   -
   \lambda \, u(x) \, v^2(x) 
   &=
   0 \>.
   \label{A.e:29b}
\end{align}
\end{subequations}

Next set
\begin{subequations}\label{S.e:3}
\begin{align}
   \tau_{u} &= \nu^{-1} \>,
   &
   \tau_{v} &= \mu^ {-1}  \>,
   \\
   l_{u} &= \sqrt{D_u \tau_{u}} \>,
   &
   l_{v} &= \sqrt{D_v \tau_{v}} \>,
\end{align}
\end{subequations}
and put
\begin{equation}\label{SC.e:tuuvdefs}
   \tu(x) = \frac{\nu}{f} \, u(x)
   \, ~~
   \tv(x) = \sqrt{\frac{\lambda}{\nu}} \, v(x) \>.
\end{equation}
Then \eqref{A.e:28} becomes
\begin{subequations}\label{A.e:30}
\begin{align}
   [\, \tau_u \partial_t - \ell_{u}^2 \, \nabla^2 + 1 \,] \, \tu(x)
   +
   \tu(x) \, \tv^2(x)
   &=
   1 \>,
   \label{A.e:30a}  \\
   [\, \tau_v \partial_t - \ell_{v}^2 \, \nabla^2 + \mu \,] \, \tv(x)
   -
   \gamma \, \tu(x) \, \tv^2(x) 
   &=
   0 \>,
   \label{A.e:30b}
\end{align}
\end{subequations}
where
\begin{equation}\label{SC.e:gammadef}
   \gamma
   =
   \frac{f}{\mu} \sqrt{\frac{\lambda}{\nu}} \>.
\end{equation}
Defining the ratios,
\begin{equation}\label{S.e:4}
   \tilde{\tau} 
   = 
   \frac{\tau_{v}}{\tau_{u}} = \frac{\nu}{\mu}
   \, ~~
   \beta 
   = 
   \frac{D_u}{D_v}
   \, ~~
   \tilde{\ell} 
   = 
   \frac{\ell_{v}}{\ell_{u}} = \sqrt{\frac{\tilde{\tau}}{\beta}} \>,
\end{equation}
and setting $\tilde{t} = \tau_u t$ and $\tilde{x} = \ell_u x$, Eqs.~\eqref{A.e:30} become:
\begin{subequations}\label{S.e:5}
\begin{align}
   [\, 
      \partial_{\tilde{t}} 
      - 
      \nabla^2_{\tilde{x}} 
      + 
      1 \, 
   ] \, \tu(\tilde{x})
   + 
   \tu(\tilde{x}) \, \tv^2(\tilde{x})
   &=
   1 \>,
   \label{S.e:5a} \\
   [\, 
      \tilde{\tau} \, \partial_{\tilde{t}} 
      -  
      \tilde{\ell}^2 \, \nabla^2_{\tilde{x}} 
      + 
      1 \, 
   ]\, \tv(\tilde{x}) 
   - 
   \gamma \,
   \tu(\tilde{x}) \, \tv^2(\tilde{x})
   &=
   0 \>,
   \label{S.e:5b}
\end{align}
\end{subequations}
which are Eqs.~(2.13) and (2.14) of Muratov and Osipov.

%
%
\subsubsection{One dimensional single spot solution}

In arbitrary spatial  dimension $d$ there exist approximate analytical forms for the  single spot solution which is a spike at the origin  of $v$, which then goes to zero out side the spike region. $u$ is non-zero inside the spike and relaxes to its constant value outside the spike.
These approximate solutions are found by a ``singular'' perturbation theory and require matching outer and inner solutions.

Eqs.~\eqref{S.e:5} in one-dimension becomes:
\begin{subequations}\label{S.f:5}
\begin{align}
   [\, \partial_{\tilde{t}} - \partial_{\tilde{x}}^2 + 1 \, ]\, \tilde{u} 
   + 
   \tilde{u} \, \tilde{v}^2
   &=
   1 \>,
   \label{S.f:5a} \\
   [\, \tilde{\tau} \, \partial_{\tilde{t}} -  \tilde{l}^2 \, \partial_{\tilde{x}}^2 + 1 \, ]\, \tilde{v} 
   - 
   \gamma \,
   \tilde{u} \, \tilde{v}^2
   &=
   0 \>,
   \label{S.f:5b}
\end{align}
\end{subequations}
In general, the profile of different types of solutions will exist in the different regimes of the limit $\tilde{l} \ll 1$.  As an example, we provide a simple spike solution, which is accurate in the limit $\tilde{l} \simeq \gamma^2 \ll 1$.
In the steady-state Eqs.~\eqref{S.f:5} reduce to,
\begin{subequations}\label{S.e:6}
\begin{align}
   \frac{\rmd^2 \tilde{u}}{\rmd \tilde{x}^2} 
   - 
   \tilde{u} 
   - 
   \tilde{u} \, \tilde{v}^2 
   &= -1 \>,
   \label{S.e:6a} \\
   \tilde{l}^2 \, \frac{\rmd^2 \tilde{v}}{\rmd \tilde{x}^2} 
   - 
   \tilde{v} 
   + 
   \gamma \, \tilde{u} \, \tilde{v}^2 
   &= 0 \>,
   \label{S.e:6b}
\end{align}
\end{subequations}
Since $u$ varies on the order of unity, and $v$ varies on the order of $\tilde{l} \ll 1$, one can separate scales inside and outside the profile.  Assuming that within the spike $\tilde{u} = \tilde{u}(0)$ is roughly a constant, substituting this into \eqref{S.e:6b} and solving for $\tilde{v}(\tilde{x})$ we get,
\begin{equation}\label{S.e:7}
   \tilde{v}(\tilde{x})
   =
   \tilde{v}(0) \, \sech^{2} \Bigl [ \frac{\tilde{x}}{2 \, \tilde{l} } \Bigr ] \>,
   \qquad
   \tilde{v}(0)
   =
   \frac{3}{2 \tilde{u}(0) \gamma} \>,
\end{equation}
where $\tilde{v}(0)$ is the amplitude of the spike.  
Away from the spike, $\tilde{v} = 0$ since this is in the background of the ``red-state'' where $v = 0, u = 1$. 
In Eq.~\eqref{S.e:6a}, the term $\tilde{v}^2(\tilde{x})$ acts as a $\delta$-function, since
\begin{equation}
\int_{-\infty}^{\infty} \sech^4 [\frac{x}{2 \epsilon}] =  \frac{8 \epsilon} {3} 
\end{equation} 
\begin{equation}\label{S.e:7.1}
   \tilde{v}^2(\tilde{x})
   \rightarrow
   B \, \delta(\tilde{x}) \>,
   \qquad
   B
   =
   \frac{6 \, \tilde{l}}{\tilde{u}^2(0) \gamma^2  } \>.
\end{equation}
Substituting \eqref{S.e:7.1} into \eqref{S.e:6a}, we get
\begin{equation}\label{S.e:7.2}
   \frac{\rmd^2 \tilde{u}}{\rmd \tilde{x}^2} 
   - 
   \tilde{u}
   +
   1
   = 
   \frac{6 \, \tilde{l}}{ \tilde{u}(0) \gamma^2} \, \delta(\tilde{x}) \>,
\end{equation}
the solution of which is
\begin{equation}\label{S.e:8}
   \tilde{u}(\tilde{x})
   =
   1 - \frac{ 3 \, \tilde{l}}{\tilde{u}(0) \gamma^2} \, \rme^{ - |\, \tilde{x} \,| } \>.
\end{equation}

Evaluating \eqref{S.e:8} at $\tilde{x} = 0$, gives
\begin{align}\label{S.e:9}
   \tilde{u}(0)
   &=
   \frac{1}{2} \,  
   \Bigl [\, 1 \pm \sqrt{1 - (\gamma_c/\gamma)^2} \, \Bigr ]
   =
   \frac{\gamma_c}{2 \gamma} \, \rme^{\pm \xi}
   \\
   &=
   \frac{\rme^{\pm \xi}}{2 \cosh{\xi}} \>.
   \notag
\end{align}
where $\gamma_c = \sqrt{12 \, \tilde{l}}$, and we have put $\gamma = \gamma_c \, \cosh{\xi}$.  From \eqref{S.e:7}, we then find
\begin{equation}\label{S.e:10}
   \tilde{v}(0)
   =
   \frac{3}{\gamma_c} \,  \rme^{\mp \xi} \>.
\end{equation}

Here one would choose the negative sign for the stable case.  The above discussion only applies for a single spot at the ``origin.'' 
In one dimension in the spot formation regime, one spot will go through a sequence of bifurcations in time until a steady state is reached.   This is seen in the simulations of Reynolds et.~al~\cite{Reynolds}. 

%
%
\subsubsection{Two dimensional single spot solution} 

For our simulations, there are two spatial dimensions and the single  spike solution  at the origin is radially symmetric.  The Laplacian in two dimensions is given by
\begin{equation}
\frac{\partial^2} {\partial r^2}  + \frac{1}{r} \frac{\partial}{\partial r} + \frac{1}{r^2} \frac{ \partial^2}{\partial \theta^2} 
\end{equation}
In two dimensions the equations \eqref{S.e:5} for radial solutions and in the steady-state reduce to
\begin{subequations}\label{S.f:6}
\begin{align}
 \frac{\partial^2 u} {\partial r^2}  + \frac{1}{r} \frac{\partial u}{\partial r}
   - 
   \tilde{u} 
   - 
   \tilde{u} \, \tilde{v}^2 
   &= -1 \>,
   \label{S.f:6a} \\
   \tilde{l}^2 \,
   \Bigl [\,
      \frac{\partial^2 v} {\partial r^2}  
      + 
      \frac{1}{r} \frac{\partial v }{\partial r} \,
   \Bigr ]
   - 
   \tilde{v} 
   + 
   \gamma \, \tilde{u} \, \tilde{v}^2 
   &= 0 \>,
   \label{S.f:6b}
\end{align}
\end{subequations}
Since $u$ varies on the order of unity, and $v$ varies on the order of $\tilde{l} \ll 1$, one can separate scales inside and outside the profile.  Assuming that within the spike $\tilde{u} = \tilde{u}(0)$ is roughly a constant, substituting this into \eqref{S.f:6b}  one needs to solve the equation: 
\begin{equation}\label{approxv}
   \tilde{l}^2 \,
   \Bigl [\,
     \frac{\partial^2 v} {\partial r^2}  
     + 
     \frac{1}{r} \frac{\partial v }{\partial r}\,
   \Bigr ]
   - 
   \tilde{v} 
   + 
   \gamma \, \tilde{u_0} \, \tilde{v}^2 
   = 0 \>.
\end{equation}

A trial wave function which is constant at the origin and has the correct falloff at large $r$ is the solution of  Eq.~\eqref{approxv}, which ignores the $(1/r) \, \partial/ \partial r$ term.
So we will assume as a first approximation: 
\begin{equation}\label{spike2d}
   \tilde{v}(\tilde{r})
   =
   \tilde{v}(0) \, \sech^{2} \Bigl [ \frac{\tilde{r}}{2 \, \tilde{l} } \Bigr ] \>,
\end{equation}
where $\tilde{v}(0)$ is the amplitude of the spike.  

Following the ideas of Osipov, we can relate in this approximation $u_0$ and $v_0$ by assuming that this trial solution satisfies Eq.~\eqref{approxv} on the ``average'' by integrating the equation over two dimensions.  This gives
\bq
v_0=  \frac{3 \log (4)}{\gamma u_0 (\log (16)-1)} \approx \frac{2.34622}{\gamma u_0}
\eq
Now when $\tl <<1$,  $v^2(r) $  is proportional to an approximation to the two dimensional 
$\delta$ function.  In the absence of angular momentum
$\delta^2 (x) = \delta(r)/(2 \pi r)$.  Noting that 
\bq
\int 2 \pi r dr ~\sech^4(r/\epsilon ) 
= 
\frac{1}{3} \pi  \epsilon ^2 (\log (16)-1) \>,
\eq
we obtain
\bq
   \sech^4(r/\epsilon )
   \approx  
   \frac{1}{3} \pi \epsilon^2 \, [\, \log (16)-1 \,] \, \frac{\delta(r)}{2 \pi r} \>.
\eq
The equation for $\tu(r)$ is
\begin{align} \label{equ} 
 \frac{\partial^2 \tu} {\partial r^2}  + \frac{1}{r} \frac{\partial \tu}{\partial r}
   - 
   \tilde{u} 
   - 
   \tilde{u} \, \tilde{v}^2 +1
   =
   0 \>. 
\end{align}

Outside the spike $v=0$, and letting $\tu(r) = 1 - a f(r)$, outside the spike we have
\bq
   \frac{d^2 f}{dr^2} +\frac{1}{r} \frac{df}{dr} - f(r) = 0 \>,
\eq
whose solution is $f(r) = K_0(r)$.  So that at large r we get a different asymptotic behavior from the one dimensional case.  Thus outside the spike
\bq
 \tu(r) = 1- a K_0(r) \>,
 \eq
 and at large $r$, 
 \bq
 K_0(r) \rightarrow  \sqrt{ \frac{\pi}{2r}} e^{-r} 
 \eq
Note that we cannot use this expression as $r \rightarrow 0$ since we assume that $\tu=\tu(0)$ inside the spike.  So the radius of the spike is approximately 
$r_s= \sqrt{ \expect{r^2}} $ , where
\bq
   \expect{r^2} 
   =
   \int r^2  \tv^2(r) 2 \pi r \dd{r}
   \Big / 
   \int \tv^2(r) 2 \pi r \dd{r} \>.
\eq
Using \eqref{spike2d} we find:
\bq
   r_s 
   = 
   \tl \,
   \sqrt{\frac{6 \, [\, 3 \zeta (3)-\log (16)\,]}{\log (16)-1}} \approx  1.67975 ~ \tl \>.
\eq
We can write the exact equation for $f(r)$ as follows:
\bq
   \frac{d^2 f}{dr^2} 
   +
   \frac{1}{r} \frac{df}{dr} 
   - 
   f(r) 
   = 
   -[\, 1-f(r) \,] \, v^2 \>.
\eq
Using the 2D Delta-function approximation for $v^2 (r)$, this becomes
\bq
   \frac{d^2 f}{dr^2} +\frac{1}{r} \frac{df}{dr} - f(r) 
   = 
   - 
   \frac{4 \pi}{3} u_0  v_0^2 \, \tl^2 \, [\log(16) - 1] \, \frac{\delta(r)}{2\pi r} \>. 
\eq
This is similar to the green function equation for the modified 2D Helmholtz equation, which is given by
\bq
   G(\mathbf{r}_1 - \mathbf{r}_2) 
   =
   \frac{1}{2 \pi} K_0(\mathbf{r}_1 - \mathbf{r}_2) \>,
\eq
so that in the delta function approximation we find:
\bq 
   u(r) = 1 - \frac{2 \, \tl^2}{3} u_0 v_0^2 \, [\log(16) -1] \, K_0(r) \>.
\eq
Using this method we obtain
\bq
 a =  6 \, \tl^2 \frac{ [\, \log(4) \,]^2} {\gamma^2 \, u_0 [\log(16) -1]} \>,
\eq
where $\gamma$ is the Euler-Mascheroni constant.  
This again does not allow one to go to $r=0$ since we replaced the $\sech^4(r)$, which is well behaved at the origin, with the delta function. 

\begin{figure*}[t]
   \centering
   \subfigure[\ $u(r)$]
   { \label{f:fig2a} \includegraphics[width=0.85\columnwidth]{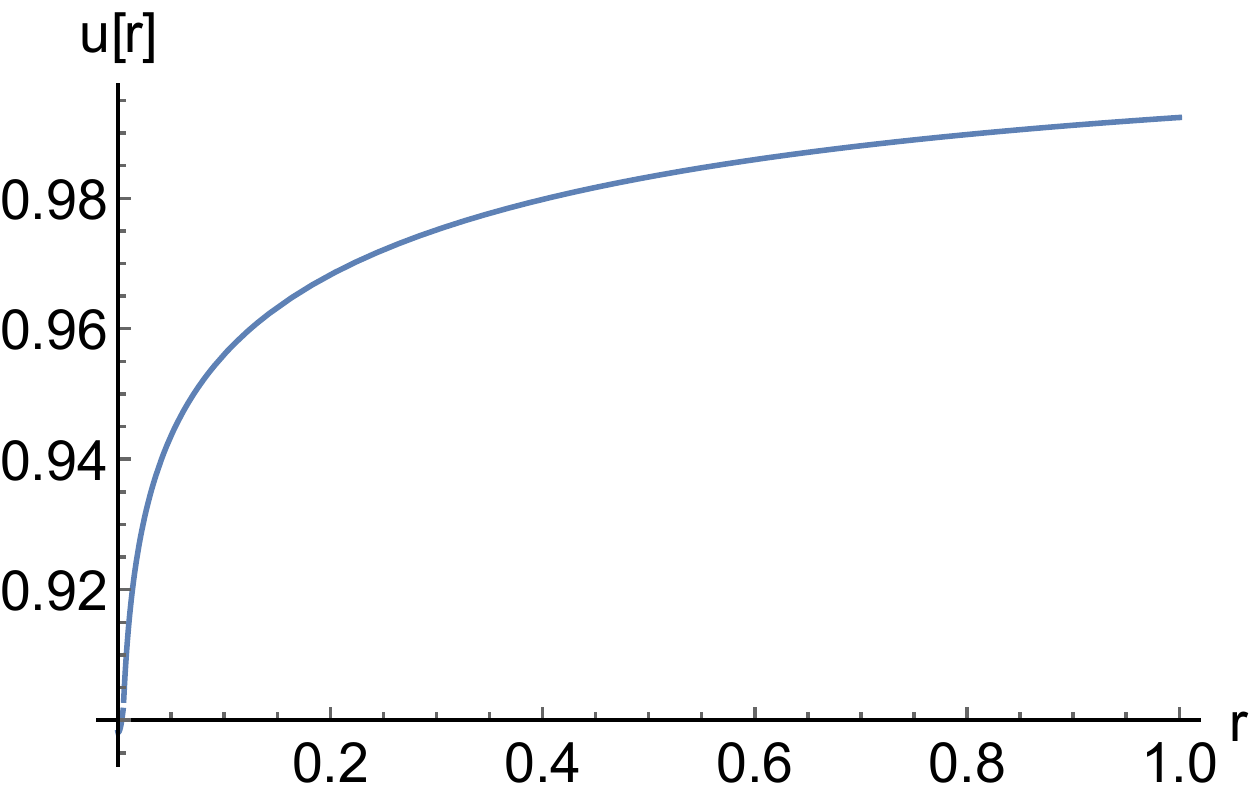} }
   \qquad
   \subfigure[\ $v(r)$]
   { \label{f:fig2b} \includegraphics[width=0.85\columnwidth]{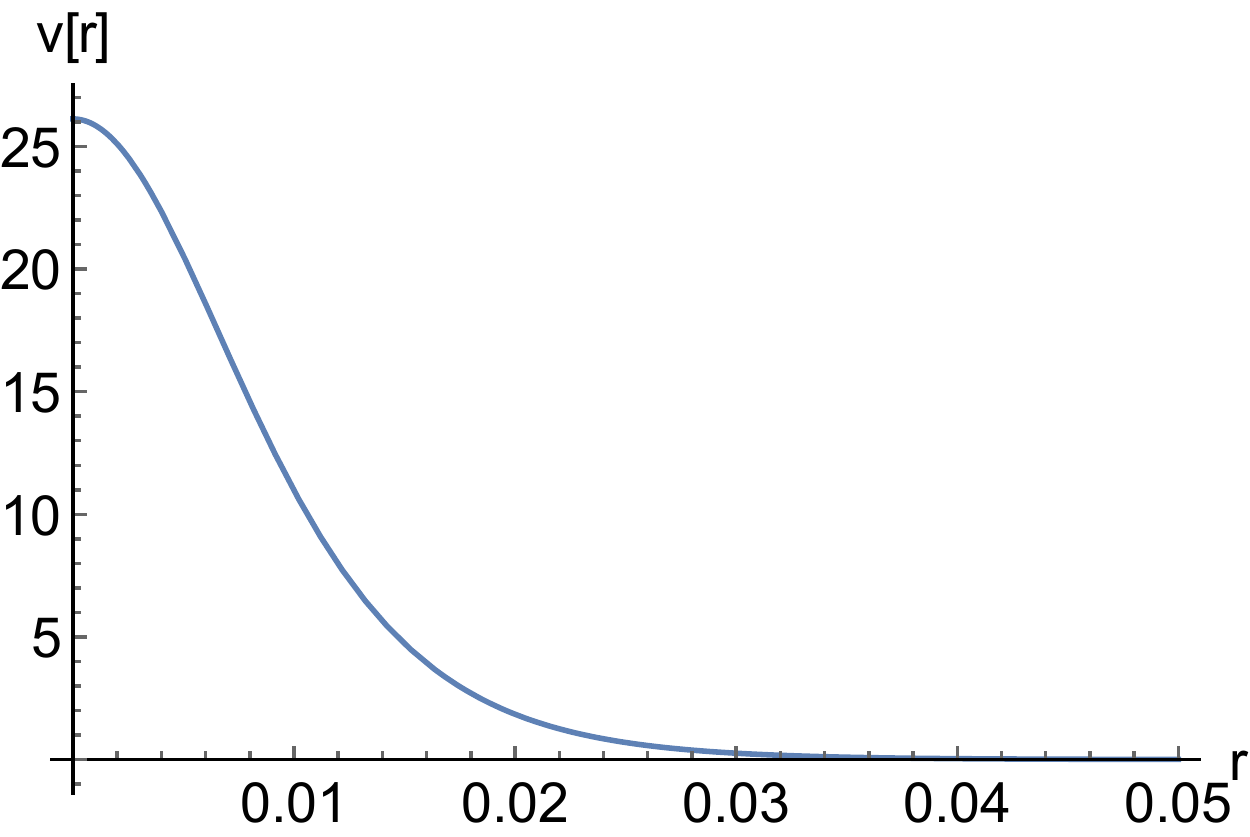} }
   \caption{$u(r)$ and $v(r)$ for spike formation when $\tl=.0054, \gamma=1/10$.}
\end{figure*}

When $r$ is of order $\tl$ we have expanding the expression for $u(r) = 1- a K_0(r)$, we find
\bq
  u(r) = a \log r+\gamma  a-a \log (2)+1 +O \left(r^2 \right) \>.
\eq
Rewriting this in terms of the dimensionless parameter $r/l$ we have 
\bq
  u(r) = 1- a \log(1/ \tl) + a \log (r/\tl )+\gamma  a-a \log (2) \>,
\eq
so that this expression is valid near $r/\tl=1$ only if  
\bq \label{caveat} 
  a \log(1/\tl) \leq 1 \>.
\eq
The constant term is numerically small:
\bq
\gamma  a-a \log (2)= -0.115932 a \>.
\eq
Let us now look at the series solution of the two differential equation near $r=0$ to see what we can learn.  First, $u(r)$ and $v(r)$ are functions of $r^2$ so that the solutions are non-singular at $r=0$.  If we are interested in the behavior in the vicinity of the spike it is good to scale $r$ by $\tl$.  In what follows we suppress the tildes.  Let $\xi = \tl/r$, which gives
\ba
   \frac{\partial^2 v} {\partial \xi^2} + \frac{1}{\xi } \frac{\partial v }{\partial \xi }\,
   - 
   \tilde{v} 
   + 
   \gamma \, \tilde{u} \, \tilde{v}^2 
   &=& 0 \>, \\
   \frac{\partial^2 u} {\partial \xi^2}  + \frac{1}{\xi } \frac{\partial u}{\partial \xi}
   - 
   \tl^2( \tilde{u} 
   +
   \tilde{u} \, \tilde{v}^2 )
   +\tl^2 
   &=& 0 \>.
   \notag
\ea
Again letting $u= 1 - f(\xi)$ we obtain
\bq 
   \frac{\partial^2 f} {\partial \xi^2}  
   + 
   \frac{1}{\xi } \frac{\partial f}{\partial \xi}
   - 
   \tl^2 f 
   + 
   \tl^2 (1-f) \, v^2 = 0 \>.
\eq
The solution of these equations up to $r^2$ are:
\ba \label{inner} 
   u_{\text{in}}(r) 
   &=&  
   u_0 
   -
   \frac{r^2}{4} \, (\, 1-u_0 -u_0 v_0^2 \,) 
   \\
   &=& u_0 -\frac{r^2}{4} \,
   \Bigl ( 1-u_0 -\frac{5.50475}{\gamma ^2 u_0} \Bigr )
   \notag \\
   v_{\text{in}}(r) 
   &=&
   v_0 + \frac{r^2}{4} \, (\, v_0-\gamma  u_0 v_0^2 \,)
   \notag
\ea
We want to match this with the outer solution at $r=\tl$, 
\bq \label{outer}
u_{\text{out}}(r) = 1- a(u_0) K_0(r)  
\eq 
Setting Eq.~\eqref{inner} = Eq.~\eqref{outer} at $r=\tl $, with the caveat \eqref{caveat}, gives an equation for $u_0$ in terms of the other parameters.
For example, if we choose $\gamma=1/10$ and  $\tl=.005$, we find that $v_0= 23.4622/u_0$ and $a=0.0162628/u_0$.  For this example, we find by matching the outer and inner solutions at $r=\tl$ that $u_0= 0.898132$ and $u(\tl)  = 0.901962$.  Combining the inner and outer solutions for $u(\tl)$, we obtain the results shown in Fig.~\ref{f:fig2a}.  For the approximate spike we obtain the results for $v(r)$ shown in Fig.~\ref{f:fig2b}.
In our 2D simulations, we seed the initial fluctuations so that we start with a small region being excited.  In our simulations the spots in the small region keep bifurcating until eventually the whole two dimensional grid size gets full of spots and a steady state is reached. 
%
%
%
%



%
%
\section{\label{s:decoupling}Decoupling}

As pointed out in Section~\ref{s:action}, the field equations with noise \eqref{A.e:25} contain a constraint field $w(x)$ which enter the noise terms and which acts similar to the composite fields found in Fermi's local version of the theory of weak interactions, except that it is made of trilinear fields. Here we want to show that when we promote the field $w(x)$ to a dynamical field by adding a diffusion term, the late time dynamics produces the \emph{same} pattern formation structure as the original Gray-Scott model.  In a previous paper \cite{PhysRevLett.111.044101}, we assumed the the relevant composites were $UV$ and $V^2$, which did not agree with the results of deriving the Gray-Scott model from the master equation.  Instead the composite $UV^2$ played the role of the intermediate boson in the interactions induced by the fluctuations.
 
We start from the equations \eqref{A.f:27}
\begin{subequations}
\begin{align}  \label{a}
   [\, \partial_t - D_{u} \, \nabla^2 + \nu \,] \, u(x)
   +
  \frac{1}{2} w(x) 
   &=
   f \>,
 \\
   [\, \partial_t - D_{v} \, \nabla^2 + \mu \,] \, v(x)
   -
 \frac{1}{2} w(x) 
   &=
   0 \>, \label{b} 
 \\
   w(x) - 2 \, \lambda \, u(x) \, v^2(x)
   &=
   0 \>,
   \label{c}
\end{align}
\end{subequations}

We promote the constraint field $w(x)$ to a dynamic field by introducing a dynamic term with diffusion coefficient $D_w$ and mass $M$ into \eqref{c}
\begin{eqnarray}
 \frac{1}{M}  [ \partial_t - D_{w} \, \nabla^2  ]  w
 +(w  - 
   2 \, \lambda \,u \, v^2 )
   &&=
   0 \>,
   \label{D3.e:1c}
\end{eqnarray} 

%
The steady state solutions of Eqs.~\eqref{a}, \eqref{b}, and \eqref{D3.e:1c} are the \emph{same} as for the Gray-Scott model.
We expect that when $D_w$ is much smaller than $D_u$ and $D_v$, the time scale for the $w(x)$ dynamics is faster than for the other chemical reactions.  We indeed find in our numerical simulations that when  $M \geq 2$,  $w(x) \rightarrow \lambda \, u(x) \, v^2(x)$ at fast time scale, and the 
evolution of pattern formation at later times is quite similar to the original Gray-Scott model evolution.

%
%
\section{Numerical Simulations}

To find the appropriate diffusion and mass parameters for the composite field $w(x)$, we have performed numerical simulation on the modified Gray-Scott model of Eqs.~\eqref{a}, \eqref{b}, and \eqref{D3.e:1c}, and compared them to the simulation of the original Gray-Scott model \eqref{A.e:27}, in several different parameter ranges where different patterns are known to occur in the original Gray-Scott model.  

%
%
\begin{table}[t]
\caption{\label{tab:parameters}Parameters for the Gray-Scott model for the points $a$ and $b$
shown in Fig.~\ref{fig:f-kappa-turing}.}
\begin{ruledtabular}
\begin{tabular}{cddddd}
& 
\multicolumn{1}{c}{$\kappa$} & 
\multicolumn{1}{c}{$f$} & 
\multicolumn{1}{c}{$D_u$} &
\multicolumn{1}{c}{$D_v$} & 
\multicolumn{1}{c}{$D_w$}  \\
\hline
a & 0.06  & 0.05 & 0.00003 & 0.00001 & 0.000005 \Tstrut \\[1pt]
b & 0.065 & 0.04 & 0.00003 & 0.00001 & 0.000005
\end{tabular}
\end{ruledtabular}
\end{table}
%
%

Two dimensional simulations were performed in a square region of total size $L_x = L_y = 2$ divided into a grid of $N \times N$ with $N = 256$.  The size of the time step was $\Delta t = 0.5$ with $30,000$ time steps per run, or until the patterns formed were stable.  Testing with values of $N = 512$ produced essentially the same results.  We used units such that $\lambda = 1$ and set $\mu = \nu + \kappa$ and $\nu = f$ in our calculations.  Values of the model parameters $(\kappa,f,D_u,D_v,D_w)$ used for formation of ridges and spots are given in Table~\ref{tab:parameters}.  We started at $t=0$ with a distorted gaussian perturbation of the \emph{red} solutions near the origin, given by
\begin{align}\label{NS.e:1}
   u(x,y,0) &= 1 - g_u(x,y) \>, \\
   v(x,y,0) &= g_v(x,y) \>, \\
   w(x,y,0) &= \lambda \, u(x,y,0) \, v^2(x,y,0) \>.
\end{align}
where
\begin{equation}\label{NS.e:2}
   g_{\alpha}(x,y) 
   = \exp 
   \Bigl \{ 
      - 
      \frac{1}{2} \Bigr [ \frac{x - x_{\alpha}}{w} \Bigr ]^2 
      - 
      \frac{1}{2} \Bigr [ \frac{y - y_{\alpha}}{w} \Bigr ]^2 \,
   \Bigr \} \>,
\end{equation}
with the offsets:
\begin{equation*}
   (x_u,y_u) 
   = 
   (-0.05,-0.02) \>,
   \quad
   (x_v,y_v) 
   = 
   (0.05,0.02) \>.
\end{equation*}
The width $w = 0.08$ was the the same for both initial fields.  Patterns usually moved from the origin out to the edges of the square region after about $10,000$ steps.  

We used two types of simulation methods.  The first was a standard split-operator or Strang splitting method, modeled after codes by Michael Quell \cite{Quell:2014aa} and used a simple fixed point iteration to solve the nonlinear part of the system.  We also used the Runge-Kutta fourth order ETDRK4 algorithm scheme of Cox and Matthews \cite{COX2002430} written in Fortran 90 but patterned after MatLab codes by Kassam \cite{r:Kassam:2003vn,r:Kassam:2005wy}.  In both codes, we used aliasing with a 2/3 rule, and enforced real densities.  We made sure that both methods produced the same results.  The MatLab program proved to be sufficient in all cases.  

We have compared at late times the modified Gray-Scott model with the Gray-Scott model for different values of the coupling between $a$ and $b$ which lead to different patterns.  After 30,000 iterations one finds the results shown in Video~\ref{vid:mov1}.  The top row is results for the Gray-Scott model and the bottom row for the modified Gray-Scott model with $D_w = 0.000005$ and $M=2$.  The results are quite similar.  The differences can be accounted for by the slightly different initial conditions in the two cases.

%
%
\begin{video*}
\href{https://www.dropbox.com/s/pcjxc7a7okznlpl/GSridges30000.mov?dl=0}
{\includegraphics[width=0.8\columnwidth]{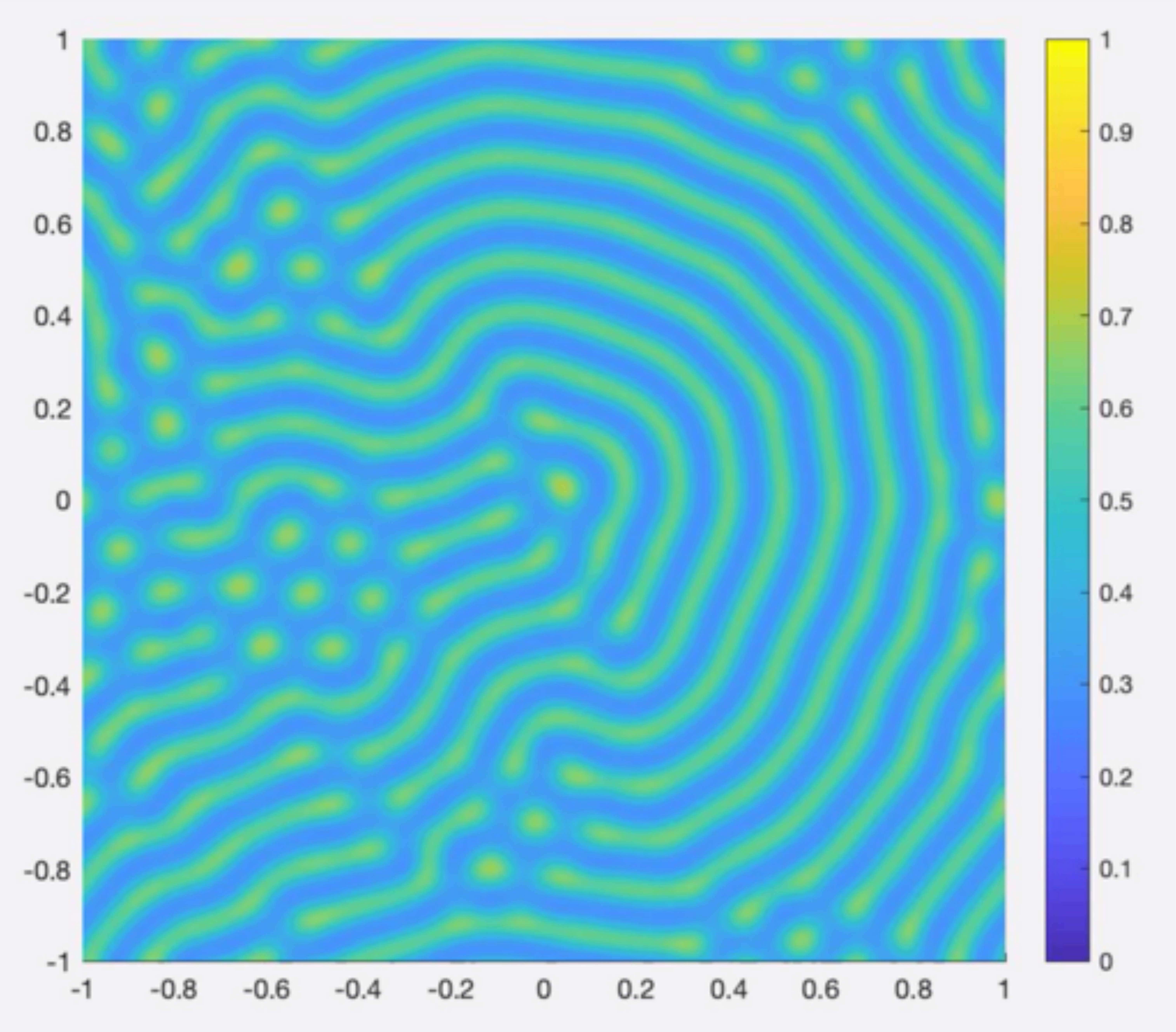}}
\qquad
\href{https://www.dropbox.com/s/qbc787konsi2amc/GSspots30000.mov?dl=0}{\includegraphics[width=0.8\columnwidth]{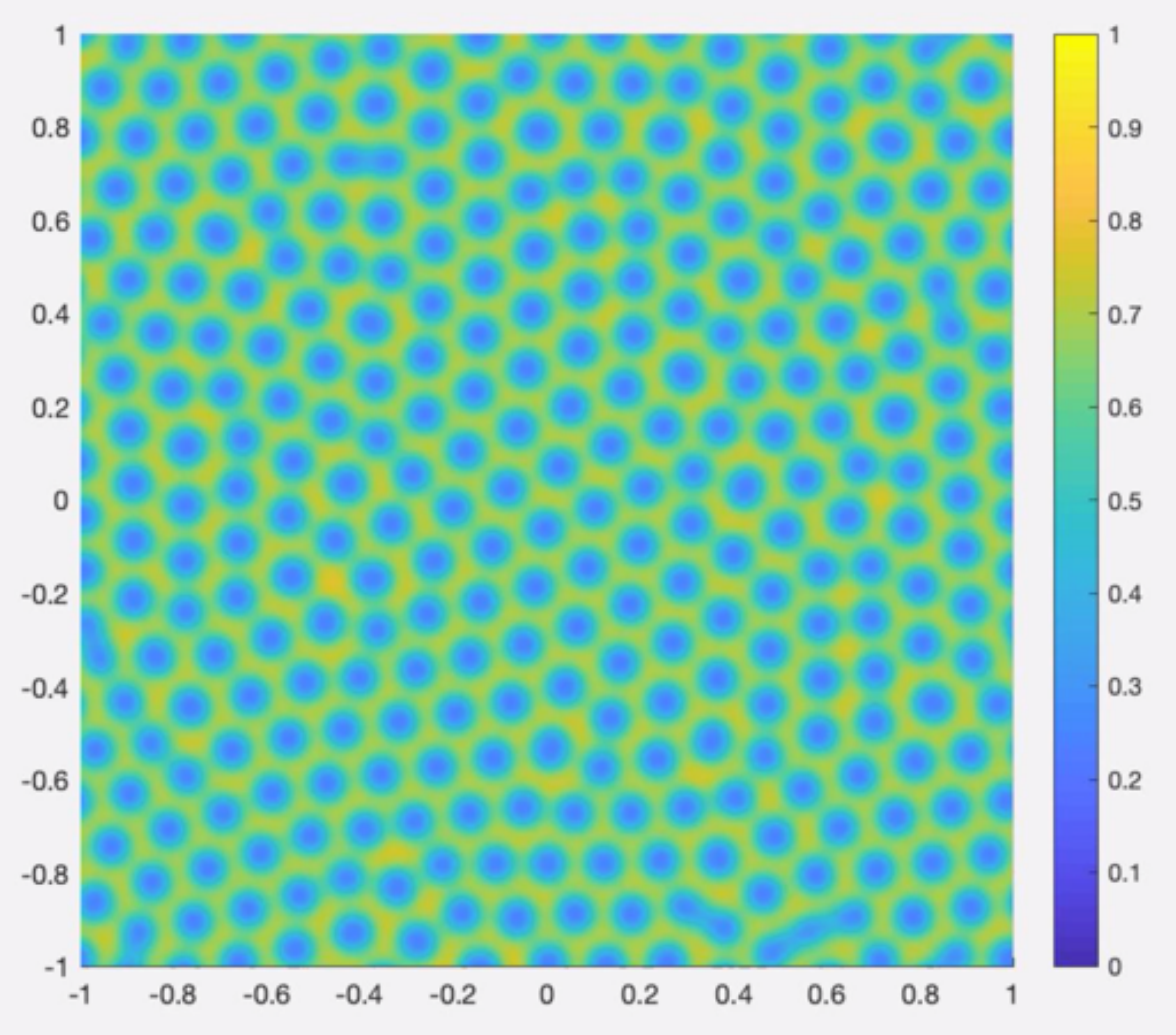}}
\\
\bigskip
\href{https://www.dropbox.com/s/wcnx1supbi4wck6/MGSridges30000.mov?dl=0}{\includegraphics[width=0.8\columnwidth]{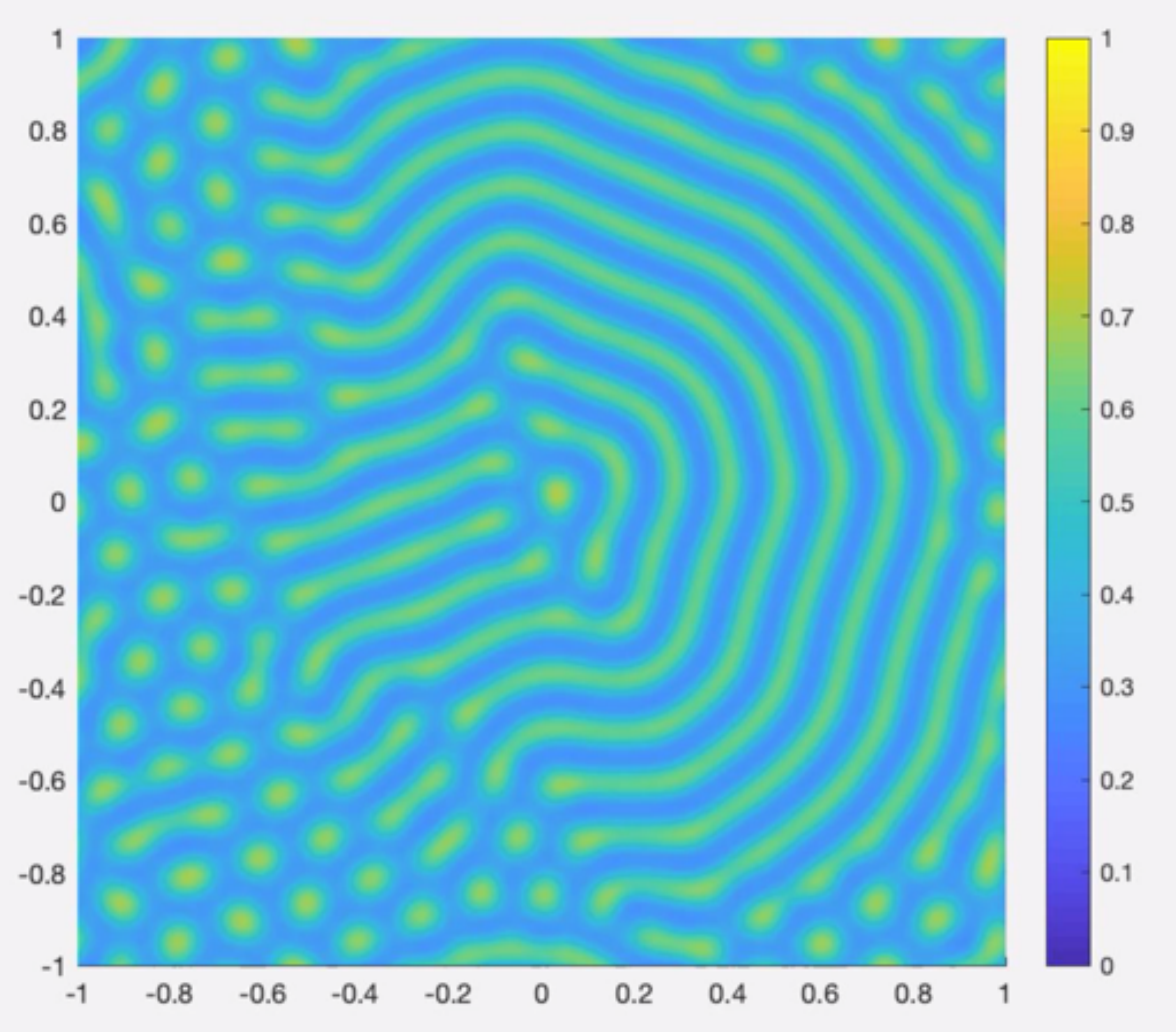}}
\qquad
\href{https://www.dropbox.com/s/47uj1g2ktpj4fxt/MGSspots30000.mov?dl=0}{\includegraphics[width=0.8\columnwidth]{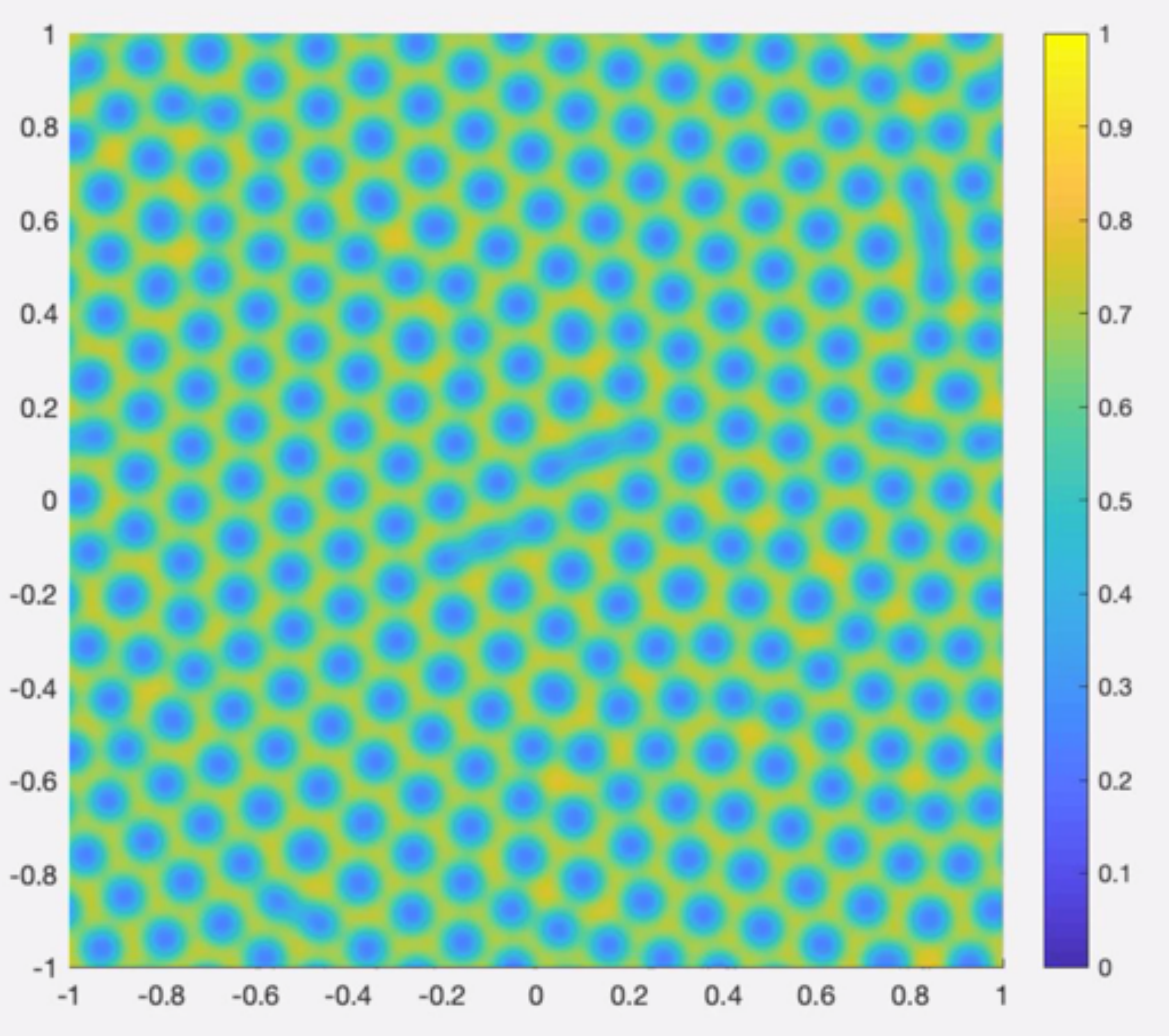}}
\caption{\label{vid:mov1}(Click on graphics to get video) 
   Top row: ridge (left) and spot (right) formation 
   for the Gray-Scott model after 30,000 iterations.
   Bottom row: ridge (left) and spot (right) formation 
   for the modified Gray-Scott model after 30,000 iterations. }
\end{video*}
%
%

%
%
\section{conclusions}

The purpose of this paper was to show that the pattern formation regions of the Gray-Scott model at late times can be reproduced by a theory with an extra fundamental composite molecule $W$ with an appropriate diffusion constant.  The composite molecule $W$ enters into the full dynamics of the chemistry of $U$ and $V$ when we consider intrinsic fluctuations. In that paper \cite{PhysRevE.88.042926} it was shown that fluctuation induced chemical reactions proceed through the intermediary of the propagation of the composite molecule $W$, so that  composite molecule $W$ is indeed the analogue of the intermediate boson of the theory of weak interactions.   

Promoting this composite molecule to having similar reaction diffusion dynamics to $U$ and $V$, makes the theory with internal noise symmetric with respect to $U$, $V$, and $W$ and provides a better way of understanding the meaning of $W$ as being driven by internal noise, since it is now considered on equal footing with $U$ and $V$.  The effect of the internal multiplicative noise on pattern formation is the subject of an ongoing investigation.
%
%
\begin{acknowledgments}
We would like to acknowledge the Santa Fe Institute for its hospitality. 
\end{acknowledgments}
%
%
\bibliography{johns.bib}
%
%
\end{document}